# Breit-Wigner distribution, quantum beats and GSI Anomaly

M. S. Hosseini and S. A. Alavi*

*Department of Physics, Hakim Sabzevari University (HSU), 96179-76487, Sabzevar, Iran*

The relationship between Breit-Wigner distribution as an underlying basis for decaying unstable quantum systems and GSI experiment (anomaly) has not been addressed properly in the literatures. We show that quantum beats can be obtained using a superposition of two Breit-Wigner distributions. This modified distribution can explain the GSI time anomaly with quantum beats resulting from the existence of two energy levels of the decaying ion.

## I. INTRODUCTION

Non-exponential decays and relaxations have been observed and studied extensively in physics, from the viewpoint of condensed matter to atomic and nuclear physics [1-29]. A recent and really challengeable observation of non-exponential decay in nuclear physics is GSI anomaly. The GSI anomaly is the periodic modulation of the expected exponential law in the EC-decays of different highly charged ions, stored at GSI, observed by the FRS/ESR Collaboration [30,31]. Many attempts have been made to explain this observation since 2008. Some authors have proposed [30-36] that the GSI anomaly is due to the interference of the massive neutrinos which compose the final electron neutrino state, but this claim has been refuted by some other studies [37-41]. There are also some other arguments in the literature. It is proposed in [42,43], that modulation in the decay of the hydrogen-like ions arises from the coupling of rotation to the spin of electron and nuclei (Thomas precession). It is suggested in [44] that the GSI Oscillations may be related to neutrino spin precession in the static magnetic field of the storage ring (ESR). It is shown in [40], that the Oscillations effect can be explained by hypothetical internal excitations of the mother ions, namely by quantum mechanical interference effect (quantum beats of the mother ion). With the help of an analogy with a double-slit experiment, it is shown in [37] that the GSI time anomaly may be caused by quantum beats due to the existence of two coherent energy levels of the decaying ion with an extremely small energy splitting.

In this work we propose a modified Breit-Wigner distribution which leads to the correct survival probability observed in GSI experiment. This new distribution can explain the GSI time anomaly with quantum beats due to the superposition of two energy states of the decaying ion.

## II. MODIFIED BREIT-WIGNER DISTRIBUTION

The basic formulae concerning the decay law of an unstable state are as follows [45]:

$$a(t) = \langle\psi|exp(-iHt)\psi\rangle = \int_{-\infty}^{+\infty} d(E)e^{-iEt}dE,$$
$$P(t) = |a(t)|^2, \tag{1}$$

where $a(t)$ and $P(t)$ are the decay survival amplitude and the survival probability respectively. $d(E)$ is the energy distribution of the unstable state and the exponential decay is given by the Breit-Wigner distribution:

$$d(E) = \frac{1}{2\pi}\frac{\Gamma}{(E-E_1)^2+\frac{\Gamma^2}{4}}, \tag{2}$$

where $\Gamma$ is the decay width. This energy distribution is satisfied in the energy normalization condition (see Appendix A):

$$\int_{-\infty}^{\infty} d(E)\, dE = 1. \tag{3}$$

If we take the Fourier transform (Eq. (1)) of the Breit-Wigner distribution, the integral gets only the contribution from the simple pole located at $E = E_1 - \frac{i\Gamma}{2}$ which leads to:

$$a_{BW}(t) = e^{-\frac{\Gamma t}{2}}e^{-iE_1 t}\ . \tag{4}$$

So we get the usual exponential law for the survival probability:

$$P(t) = |a(t)|^2 = e^{-\Gamma t}\ . \tag{5}$$

Another fundamental relation for an exponentially decaying unstable system is:

$$N(t) = N(0)e^{-\Gamma t} = N(0)\, P(t). \tag{6}$$

Eqs. (5) and (6) lead to the following relation for the number of un-decayed systems at time $t$:

$$\frac{dN(t)}{dt} = N(0)\frac{dP}{dt} = -N(0)\Gamma e^{-\Gamma t} \propto -\Gamma e^{-\Gamma t}. \tag{7}$$

As mentioned before, in 2008, the FRS/ESR Collaboration [30] (see also [31]) observed a periodically modulated exponential $\beta$-decay law of highly charged stored ions at GSI laboratory:

* Corresponding author.
*E-mail address:* s.alavi@hsu.ac.ir (S.A. Alavi).

$$\frac{dN(t)}{dt} = N(0)\frac{dP}{dt} \propto -\,\Gamma e^{-\Gamma t}[\,1\, + a\, \cos(\omega t + \varphi)\,]. \qquad (8)$$

Many theoretical attempts have been made since then to explain this unusual periodic decay law. In this work, we propose a new energy distribution for the unstable state and show that it leads to the oscillations such as the ones observed in the GSI experiment. In fact, it is easy to understand that oscillations such as the ones observed in GSI experiment would not be obtained if we use the standard (usual) Breit-Wigner distribution, so we need to go beyond these widely used formulae.

Using Eq. (7) and the integrals given in the Appendix A, we have the following expression for the survival probability in GSI experiment:

$$P(t) \propto e^{-\Gamma t}\left\{1 + \frac{a\,\Gamma^2}{\Gamma^2+\omega^2}\cos(\omega t + \varphi) \; - \frac{a\,\omega\Gamma}{\Gamma^2+\omega^2}\sin(\omega t + \varphi)\right\}, \qquad (9)$$

which can be rewritten in the following form:

$$P(t) \propto e^{-\Gamma t}\left\{1 + \frac{a\,\Gamma^2}{\Gamma^2+\omega^2}\cos(\omega t + \varphi) + \frac{a\,\omega\Gamma}{\Gamma^2+\omega^2}\cos\left(\omega t + \varphi + \frac{\pi}{2}\right)\right\}. \qquad (10)$$

If we set $a = 0$, we recover the usual exponential decay law, i.e., Eq. (5).

Now we suppose the following form for the survival probability:

$$P(t) = \left|\int \frac{a_1\, e^{-i\,E\,t}\, dE}{(E-E_1)^2+\frac{\Gamma^2}{4}} + \int \frac{a''_1\, e^{-i\,E\,(t+\Delta t)}\, dE}{(E-E'_1)^2+\frac{\Gamma^2}{4}}\right|^2, \qquad (11)$$

which shows the interference between two amplitudes at time $t$ that are not coherent ($\Delta t$ is costant), see Appendix B for more details.

The energy distributions corresponding to the first and the second amplitudes in Eq. (11), are $\frac{\Gamma}{2\pi}\frac{a_1}{(E-E_1)^2+\frac{\Gamma^2}{4}}$ and $\frac{\Gamma}{2\pi}\frac{a''_1}{(E-E'_1)^2+\frac{\Gamma^2}{4}}$ respectively. $a_1$ and $a''_1$ are real parameters. Using Eqs. (3) and (A2), one can easily derive the normalization condition, $a_1 + a''_1 = 1$.

The solutions of the integrals are presented in Appendix A, so we have:

$$P(t) = e^{-\Gamma t}\left|a_1\, e^{-i\,E_1 t} + a'_1\;\, e^{-i\,E'_1\,(t+\Delta t)}\,\right|^2, \qquad (12)$$

where $a'_1 = a''_1 e^{-\Gamma\Delta t}$ (we suppose $\Delta t = 0$ at $t = 0$, because there is no decay and interference at $t = 0$, so $a'_1 = a''_1$, at $t = 0$). Introducing the energies $E_1$ and $E'_1$ , so that $E'_1 - E_1 = \omega$, for example as:

$$E_1 = E - \omega/2\,, \quad E'_1 = E + \omega/2. \qquad (13)$$

we arrive at the following expression:

$$P(t) = e^{-\Gamma t}\left[a_1^2 + a'^2_1 + 2\, a_1 a'_1\, \cos\left((E'_1 - E_1)t + E'_1\Delta t\right)\right]$$
$$= e^{-\Gamma t}[a_1^2 + a'^2_1 + 2\, a_1 a'_1\, \cos(\omega t + E'_1\Delta t)], \qquad (14)$$

where $E'_1\Delta t$ is a constant phase. Now we suppose there are two groups A (or 1) and B (or 2) of systems with energy levels $E_1$ and $E'_1$. If for the systems in group A, we have $E'_1\Delta t = \varphi$, then the survival probability $P_1(t)$ can be written as:

$$P_1(t) = e^{-\Gamma t}[a_1^2 + a'^2_1 + 2 a_1 a'_1 \cos(\omega t + \varphi)], \qquad (15)$$

and for the systems in group B, we suppose $E'_1\Delta t = \varphi + \frac{\pi}{2}$, so we have:

$$P_2(t) = e^{-\Gamma t}\left[a_2^2 + a'^2_2 + 2\, a_2 a'_2\, \cos\left(\omega t + \varphi + \frac{\pi}{2}\right)\right]$$
$$= e^{-\Gamma t}[a_2^2 + a'^2_2 - 2\, a_2 a'_2\, \sin(\omega t + \varphi)], \qquad (16)$$

with $a_2 + a'_2 = 1$. So the total probability reads:

$$P(t) = \frac{N_1(t)}{N(t)}P_1(t) + \frac{N_2(t)}{N(t)}P_2(t)$$
$$= e^{-\Gamma t}\left\{\frac{N_1(t)}{N(t)}[a_1^2 + a'^2_1 + 2\, a_1 a'_1 \cos(\omega t + \phi)] + \frac{N_2(t)}{N(t)}[a_2^2 + a'^2_2 - 2\, a_2 a'_2\, \sin(\omega t + \varphi)]\right\}, \qquad (17)$$

where $N(t) = N_1(t) + N_2(t)$, is the total number of un-decayed systems at time t, and $N_i(t), i = 1{,}2$, are the numbers of un-decayed systems in groups A and B, respectively. The probabilities (15)-(17), at time t = 0 (so $\Delta t = 0$), are given by:

$$P_1(0) = (a_1 + a'_1)^2 = 1$$
$$P_2(0) = (a_2 + a'_2)^2 = 1$$
$$P(0) = \frac{N_1(0)}{N(0)}P_1(0) + \frac{N_2(0)}{N(0)}P_2(0)$$
$$= \left\{\frac{N_1(0)}{N(0)}[(a_1 + a'_1)^2] + \frac{N_2(0)}{N(0)}[(a_2 + a'_2)^2]\right\}$$
$$= \frac{N_1(0)+N_2(0)}{N(0)} = 1, \qquad (18)$$

where we have used the normalization condition of the energy distributions $a_i + a'_i = 1, i = 1{,}2$.

Therefore, all the probabilities satisfy in the correct initial conditions.

## III. QUANTUM BEATS AND MODIFIED BREIT-WIGNER FORMULA

The quantum beat spectroscopy is a universal tool for the determination of the excited state structures in gases, liquids, and solid states. According to the time-energy uncertainty principle $\Delta E\; \Delta t \geq \hbar$, if a limited time is used to determine the energy, one can only determine it with an uncertainty $\Delta E$. So, if several quantum levels exist within $\Delta E \ \geq \frac{\hbar}{\Delta t}$, these levels will be excited simultaneously and produce a superposition of several eigenstates resulting in a sinusoidally oscillating and exponentially decaying signal. This effect is known as quantum beats and is due to the interference between the amplitudes emitted from two or more near-degenerate eigenstates.

Mathematically, the excited state can be expressed as a linear combination of the states. For the two states system shown in Fig. 1, we have:

$$|\psi(0)\rangle = \ c_1|1\rangle + \ c_2|2\rangle. \tag{19}$$

In general the time evolution of this state is given by:

$$|\psi(t)\rangle = c_1\, e^{-iE_1t+i\varphi_1}\, |1\rangle + c_2\, e^{-iE_2t+i\varphi_2}\, |2\rangle. \tag{20}$$

See Appendix C for more explanation. So, the amplitude of the evolution of the system is:

$$\langle\psi(0)|\psi(t)\rangle = [|c_1|^2\, e^{-iE_1t} + |c_2|^2 e^{-iE_2t+i\varphi}]e^{i\varphi_1}, \tag{21}$$

where $\varphi = \varphi_2 - \varphi_1$. We note that $\varphi_1 = \varphi_2 = 0$, so $\varphi = 0$ at $t = 0$ and the normalization condition is $|c_1|^2 + |c_2|^2 = 1$. If we consider the following expressions for the energies $E_1$ and $E_2$:

$$E_1 = \ E - \frac{\omega}{2} - i\frac{\Gamma}{2}$$

$$E_2 = \ E + \frac{\omega}{2} - i\frac{\Gamma}{2}$$

So, $E_2 - E_1 = \ \omega$ and we have:

$$|\langle\psi(t)|\psi(0)\rangle|_\varphi{}^2 = \ e^{-\Gamma t}\{|c_1|^4 \ + |c_2|^4$$
$$+2\, |c_1|^2\, |c_2|^2 \cos(\omega t + \varphi)\}. \tag{22}$$

Again we suppose there are two groups A and B of systems with energy levels $E_1$ and $E_2$. If for the systems in group A, we have $\varphi_A = \ \varphi$, then the probability to describe the evolution of the system is as follows:

$$|\langle\psi(t)|\psi(0)\rangle|_\varphi^2 = \ e^{-\Gamma t}\{|c_1|^4 \ + |c_2|^4$$
$$+2\, |c_1|^2\, |c_2|^2 \ \cos(\omega t + \varphi)\}. \tag{23}$$

For the systems in group B, we suppose:

$$|\psi(0)\rangle = \ c'_1|1\rangle + \ c'_2|2\rangle, \tag{24}$$

We take $\varphi_B = \ \varphi + \frac{\pi}{2}$, so:

$$|\langle\psi(t)|\psi(0)\rangle|_{\varphi+\frac{\pi}{2}}^2 = e^{-\Gamma t}\{|c'_1|^4 \ + |c'_2|^4$$
$$+2\, |c'_1|^2\, |c'_2|^2 \ \cos(\omega t + \varphi_B)\}$$
$$= e^{-\Gamma t}\{|c'_1|^4 \ + |c'_2|^4$$
$$-2\, |c'_1|^2\, |c'_2|^2 \ \sin(\omega t + \varphi)\}. \tag{25}$$

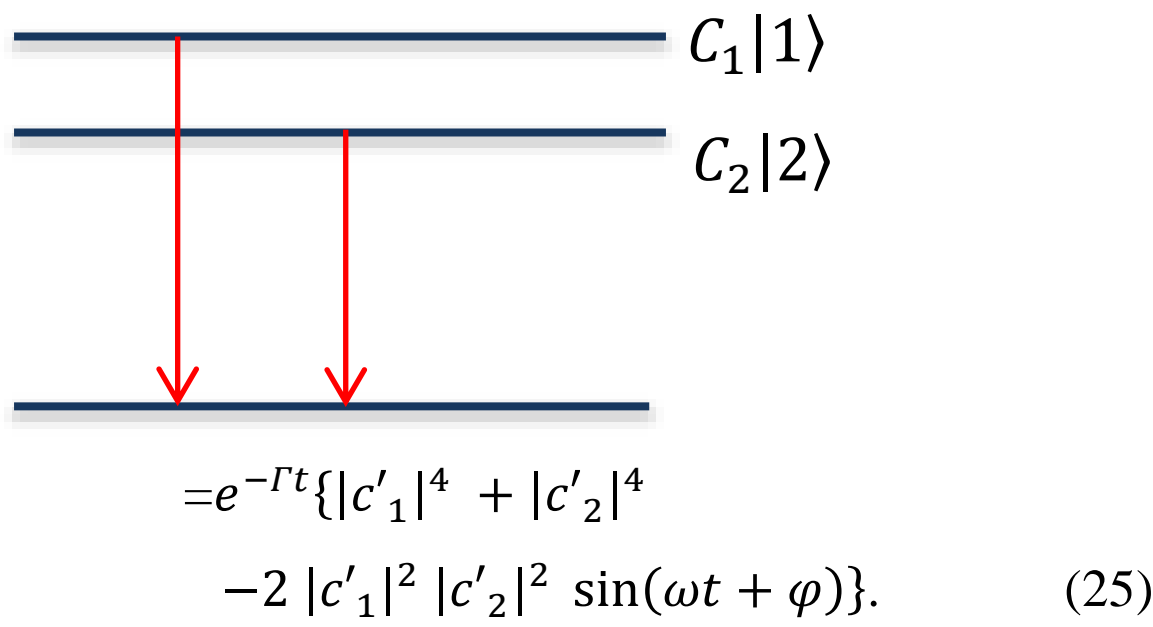

FIG. 1. Energy levels for a system with two upper and one lower state.

$\varphi_B$ is also zero at $t = 0$ and the normalization condition is $|c'_1|^2 + |c'_2|^2 = 1$. We note that $\varphi_A = 0$ and $\varphi_B = 0$ at $t = 0$. It is easy to check that $|\langle\psi(0)|\psi(0)\rangle|_{\varphi+\frac{\pi}{2}}^2 = (|c'_1|^2 + |c'_2|^2)^2 = 1$. Therefore, the total probability is given by:

$$|\langle\psi(t)|\psi(0)\rangle|^2 = \frac{N_1(t)}{N(t)}|\langle\psi(t)|\psi(0)\rangle|_\varphi^2$$
$$+\frac{N_2(t)}{N(t)}|\langle\psi(t)|\psi(0)\rangle|_{\varphi+\frac{\pi}{2}}^2$$
$$= e^{-\Gamma t}\Big\{\frac{N_1(t)}{N(t)}[|c_1|^4 \ + |c_2|^4$$
$$+2\, |c_1|^2\, |c_2|^2 \ \cos(\omega t + \varphi)]$$
$$+\frac{N_2(t)}{N(t)}[|c'_1|^4 \ + |c'_2|^4$$
$$-2\, |c'_1|^2\, |c'_2|^2 \ \sin(\omega t + \varphi)]\Big\}, \tag{26}$$

which is the same as Eq. (17), obtained using Breit-Wigner distribution. At time $t = 0$, we get:

$$|\langle\psi(0)|\psi(0)\rangle|^2 = \frac{N_1(0)}{N(0)}(|c_1|^2 + \ |c_2|^2)^2$$
$$+\frac{N_2(0)}{N(0)}(|c'_1|^2 + \ |c'_2|^2)^2$$
$$= \frac{N_1(0)+N_2(0)}{N(0)} = 1. \tag{27}$$

## IV. DISCUSSION

In this section, we answer some questions may naturally arise.

a). We have shown that by adding two Breit-Wigner amplitudes having different energies and a time dephasing between them, the survival and decay probabilities acquire a beat structure. But to which different eigenstates these different energies are alluding?. Or, which pair of states leads to the oscillations. Are they nuclear states of these hydrogen-like ions or different mass eigenstates of the neutrinos?

b). We also suggested that quantum beats are responsible for the non-purely exponential decay of these radioactive isotopes

(Eq. (22)). But what is the actual splitting $\omega$ between states 1 and 2 and what are these two states?

We start with mentioning that the Breit-Wigner distribution (Eq. (2)), is used to model resonances (unstable particles or systems) in high-energy and nuclear physics. $E$ is the energy that produces the resonance, $M$ is the mass of the resonance, and $\Gamma$ is the decay width. In other words, this distribution shows the energy distribution of the unstable system before decay not the products of the decay. So in the case of our study it is the energy distribution of the (parent nucleus + captured electron), not the products of the decay like neutrinos.

To proceed further, we introduce a more recent paper than the papers mentioned so far. In [46], the authors proposed a model in which the observed modulations arise from the coupling of rotation to the spins of electron and nucleus. They also showed that the modulations are connected to quantum beats and to the effect of the Thomas precession on the spins of bound electron and nucleus, the magnetic moment precessions of electron and nucleus and their cyclotron frequencies. The authors stressed the fact that the paper differs in essential ways from their previous work [42] and represents a more complete version of results given in [43], because it takes into account all the relevant features of the GSI experiment and can explain the different aspects of the experiment including the observed time modulation $T = 7.1\ s$.

The model also predicts that the anomaly cannot be observed if the motion of the ions is rectilinear, or if the ions are stopped in a target.

The main ingredients of their model are the Thomas precessions $\omega_{Th}^{(e,n)}$, the precessions due to the magnetic moments $\omega_{ge}$ and $\omega_{gn}$ and the angular cyclotron frequencies $\omega_c^{(e,n)}$ that contain the Mashhoon term.

The Hamiltonian of the system is [46]:

$$H_1 = -A\boldsymbol{s}.\boldsymbol{I} - \boldsymbol{s}.\boldsymbol{\Omega}_e - \boldsymbol{I}.\boldsymbol{\Omega}_n. \tag{28}$$

where $\boldsymbol{\Omega}_e$ and $\boldsymbol{\Omega}_n$ are given by:

$$\boldsymbol{\Omega}_e \equiv \omega_{ge} + \omega_{Th}^{(e)} - \omega_c^{(e)}. \tag{29}$$

$$\boldsymbol{\Omega}_n \equiv \omega_{gn} + \omega_{Th}^{(n)} - \omega_c^{(n)}. \tag{30}$$

$A$ is the strength of spin-spin coupling. But they showed that the spin-spin coupling of electron and nucleus does not contribute to the modulation because these terms average out during the time of flight of the ions, or cancel out.

Now let us consider the total angular momentum operator $\hat{F} = \hat{s} + \hat{I}$, where $s$ and $I$ are the spin of the electron and the nucleus respectively and their eigenstates are indicated by $|I, m_I\rangle_I$ and $|s, m_s\rangle_s$. Then angular momentum $F$ takes the values $F = \frac{3}{2}, \frac{1}{2}$, $m_F = \pm\frac{3}{2}, \pm\frac{1}{2}$.

By diagonalizing the $(6 \times 6)$ matrix $\langle \phi_i | \hat{H}_1 | \phi_j \rangle$ there are six eigenstates as follows:

$$\phi_1 \equiv \left|\frac{3}{2},\frac{3}{2}\right\rangle_F,\ \phi_2 \equiv \left|\frac{3}{2},\frac{1}{2}\right\rangle_F$$

$$\phi_3 \equiv \left|\frac{3}{2},-\frac{1}{2}\right\rangle_F,\ \phi_4 \equiv \left|\frac{3}{2},-\frac{3}{2}\right\rangle_F$$

$$\phi_5 \equiv \left|\frac{1}{2},\frac{1}{2}\right\rangle_F,\ \phi_6 \equiv \left|\frac{1}{2},-\frac{1}{2}\right\rangle_F$$

To apply the results to the GSI experiment, it is argued in [46] that, only the states with $F = \frac{1}{2}$ are relevant because the heavy nucleus decays via EC. So one can consider only the Hilbert subspace spanned by the states $\{|5\rangle, |6\rangle\}$ and the initial ket can be considered as follows:

$$|\psi(0)\rangle = c_5|5\rangle + c_6|6\rangle. \tag{31}$$

where [46]:

$$|5\rangle = \frac{B_-}{\sqrt{1+B_-^2}}\phi_2 + \frac{1}{\sqrt{1+B_-^2}}\phi_4, \tag{32}$$

$$|6\rangle = -\frac{A_-}{\sqrt{1+A_-^2}}\phi_3 + \frac{1}{\sqrt{1+A_-^2}}\phi_6, \tag{33}$$

and

$$A_- = \frac{9A - 2\boldsymbol{\Omega}_e + 2\boldsymbol{\Omega}_n - 3\sqrt{\Delta_-}}{4\sqrt{2}\,(\boldsymbol{\Omega}_e - \boldsymbol{\Omega}_n)}, \tag{34}$$

$$B_- = \frac{9A + 2\boldsymbol{\Omega}_e - 2\boldsymbol{\Omega}_n - 3\sqrt{\Delta_+}}{4\sqrt{2}\,(\boldsymbol{\Omega}_e - \boldsymbol{\Omega}_n)}. \tag{35}$$

So the states $|1\rangle$ and $|2\rangle$ in Eq. (19) (or states with energies $E$ and $E'$ in Eq. (11)) and the splitting $\omega$ between states $|1\rangle$ and $|2\rangle$ could be the states $|5\rangle$ and $|6\rangle$ introduced in expressions (32), (33) and $\omega_{56} = \frac{\boldsymbol{\Omega}_e - 4\,\boldsymbol{\Omega}_n}{3}$ repectively.

The situation for ions in storage ring is similar to that of muons in g-2 experiment so we briefly review the Muon's decay in g-2 experiment.

## V. MUON G-2 EXPERIMENT

The muon g-2 experiment examines the precession of muons that are subjected to a magnetic field to measure the anomalous magnetic dipole moment of a muon. The main goal is to test the Standard Model's predictions of this value.

Muons are fed into a uniform, doughnut-shaped magnetic field in a storage ring and travel in a circle. After circling the ring many times, muons spontaneously decay to electron (plus neutrinos) and the decay distribution has the following form [47]:

$$N(t) = N_0(E)e^{-\Gamma t}\{1 + A(E)\ \sin(\omega_a t + \phi(E))\}. \tag{36}$$

where $\omega_a = a_\mu + \frac{e\,B}{a_\mu}$, $B$ is the magnetic field subjected to muon and $a_\mu$ is its anomalous magnetic moment. This is an interesting expression which is analogous to Eq. (8) (see also Fig. (2)).

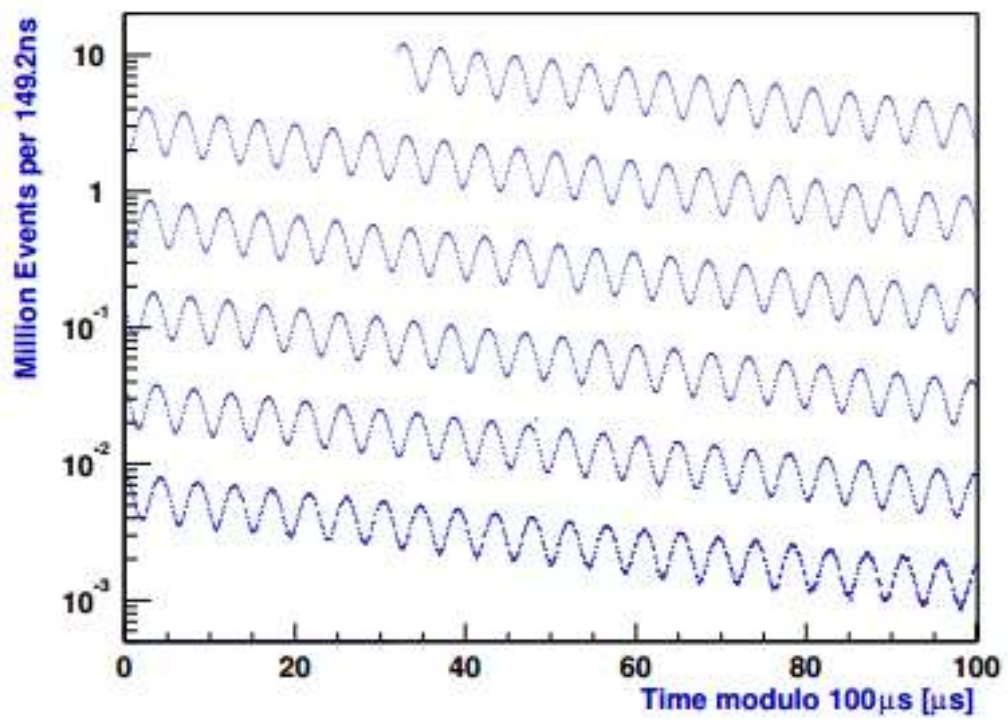

FIG. 2. Histogram with 3.6 billion decays of $\mu^-$ [47].

As mentioned before, the situation for decaying ions in GSI experiment is similar to muons in g-2 experiment and this can be a confirmation for the idea of [46] that periodic modulation in GSI experiment is due to the interaction of ions with magnetic field in storage ring. Both ions and muons are moving in circling motion and in the presence of a magnetic field. The results are also the same, there is a periodic modulations in the decay rate for both muons and ions.

So for muons or ions at rest (or rectilinear motion) or in the absence of magnetic field we get pure exponential decay (as it should) but in a storage ring (circling motion + magnetic field) there is a periodic modulation in the decay rate.

## VI. CONCLUSION

The Breit-Wigner energy distribution has a fundamental significance in the study of unstable quantum systems. In this work we have established a connection between the Breit-Wigner formula and quantum beats. We have shown that quantum beats can be obtained using superposition of Breit-Wigner distributions. This modified distribution can explain the GSI time anomaly with quantum beats resulting from the existence of two energy levels of the decaying ion.

## ACKNOWLEDGEMENTS

We would like to thank Carlo Giunti (INFN, Turin), Behnam Azadegan (HSU), Caslav Brukner (Director of the Institute for Quantum Optics and Quantum Information (IQOQI, Vienna)), Bahram Mashhoon (University of Missouri), Mohammad Khorrami (Alzahra University), Niranjan Shivaram (University of Arizona and Lawrence Berkeley National Laboratory, USA), and Pablo L. Saldanha (Universidade Federal de Minas Gerais, Brazil) for their useful discussions. S. A. Alavi would like to thank INFN, Turin, for its hospitality where some parts of this research were conducted.

## Appendix A: Some useful integrals

The useful integral to calculate the survival probability in GSI experiment is as follows:

$$\int e^{a\,x}\,\cos(b\,x)\,dx = \frac{e^{a\,x}}{a^2+b^2}\left[a\,\cos(b\,x) + b\,\sin(b\,x)\right]. \quad \text{(A1)}$$

Two helpful integrals used in this work are:

$$\int_{-\infty}^{+\infty} \frac{dE}{(E-E_1)^2 + (\Gamma/2)^2} = \frac{2\pi}{\Gamma}, \quad \text{(A2)}$$

$$\frac{1}{2\pi i}\int_{-\infty}^{+\infty} \frac{e^{-i\,E\,t}\,dE}{(E-E_1)+i\,\Gamma/2} = e^{-i(E_1 - i\,\Gamma/2)t} = e^{(-\Gamma/2 - i\,E_1)t}. \quad \text{(A3)}$$

## Appendix B: Partial Coherence- Fourier Transforms- The Convolution Integral - Autocorrelation and Cross-Correlation

Let us consider the interference of two beams at point $P$ which are partially coherence, the resulting electric field may be written as follows [48]:

$$\begin{aligned}\vec{E}_P(t) &= \vec{E}_1(t) + \vec{E}_2(t+\tau)\\ &= \vec{E}_{01}(t)e^{-i\omega t} + \vec{E}_{02}(t)e^{-i\omega(t+\tau)}.\end{aligned} \quad \text{(B1)}$$

$\tau$ is the time difference between two waves (coherence time). Then the irradiance at point $P$ is:

$$I_P = I_1 + I_2 + 2\,\Re\langle \vec{E}_1.\vec{E}_2^{\,*}\rangle. \quad \text{(B2)}$$

$I_1$ and $I_2$ represent the irradiances of the individual beams and the third term represents interference between them. One can define correlation function as:

$$\Gamma_{12}(\tau) = \langle E_1(t)\,E_2^{\,*}(t+\tau)\rangle. \quad \text{(B3)}$$

So the irradiance at $P$ may be written as:

$$I_P = I_1 + I_2 + 2\,I_1\,I_2\,\Re\,[\Gamma_{12}(\tau)]. \quad \text{(B4)}$$

Now it is useful to review the relation between Fourier transform and correlation functions. $F(\omega)$ is the Fourier transform of $f(t)$, if:

$$F(\omega) = \int_{-\infty}^{+\infty} f(t)\,e^{i\omega t}dt, \quad \text{(B5)}$$

and $f(t)$ itself is said to be the inverse Fourier transform of $F(\omega)$:

$$f(t) = \frac{1}{2\,\pi}\int_{-\infty}^{+\infty} F(\omega)\,e^{-i\omega t}d\omega. \quad \text{(B6)}$$

The so-called convolution integral which describes the convolution of two functions $f(x)$ and $h(x)$, is defined as:

$$g(X) = \int_{-\infty}^{+\infty} f(x)\, h(X-x)\, dx. \tag{B7}$$

Now, let us evaluate the following integral:

$$\int_{-\infty}^{+\infty} f(t+\tau)\, f^*(t)\, dt. \tag{B8}$$

Using inverse Fourier transform of $f^*(t)$, we have:

$$\int_{-\infty}^{+\infty} f(t+\tau)\left[\frac{1}{2\pi}\int_{-\infty}^{+\infty} F^*(\omega)\, e^{i\omega t} d\omega\right] dt, \tag{B9}$$

$$\frac{1}{2\pi}\int_{-\infty}^{+\infty} F^*(\omega)\left[\int_{-\infty}^{+\infty} f(t+\tau)\, e^{i\omega t} dt\right] d\omega, \tag{B10}$$

but:

$$f(t+\tau) = \frac{1}{2\pi}\int_{-\infty}^{+\infty} F(\omega)\, e^{-i\omega(t+\tau)} d\omega, \tag{B11}$$

So:

$$\int_{-\infty}^{+\infty} f(t+\tau)\, f^*(t)\, dt = \frac{1}{2\pi}\int_{-\infty}^{+\infty} F^*(\omega)\, F(\omega)\, e^{-i\omega\tau} d\omega. \tag{B12}$$

The left-hand side of this formula is defined as the autocorrelation of $f(t)$ and is denoted by

$$\begin{aligned} C_{ff}(\tau) &= \int_{-\infty}^{+\infty} f(t+\tau)\, f^*(t)\, dt \\ &= \int_{-\infty}^{+\infty} f(t)\, f^*(t-\tau)\, dt. \end{aligned} \tag{B13}$$

So $C_{ff}(\tau)$, is the inverse Fourier transform of $F(\omega)$, for more details see e.g. [49]. Similarly, the cross correlation of the functions $f(t)$ and $h(t)$ is defined as:

$$C_{fh}(\tau) = \int_{-\infty}^{+\infty} f^*(t)\, h(t+\tau)\, dt. \tag{B14}$$

One can replace $\tau$ by $\Delta t$.

Correlation analysis is essentially a means for comparing two signals in order to determine the degree of similarity between them.

Cross-correlation and autocorrelation are important analytic techniques for comparing sets of data. Autocorrelation represents the degree of similarity between a given set of data and a time-lagged version of that data set, and cross-correlation is a measure of similarity of two series of data as a function where the displacement of one is relative to the other. Cross- and auto-correlations are power tools which are widely used in wave optics and image processing science.

So one can interpret the probability in Eq. (12), as the correlation between two amplitudes at time $t$.

## Appendix C: Phase difference between different energy states due to some interactions

Interference phenomena on quantum systems can lead to extremely curious physics and are still in the heart of quantum mechanics. It is well known that potential differences (electric, magnetic or gravitational), are detectable by interference experiments and are of high physical significance. If a beam of particles is split into two parts, passing through two force free paths but with a definite potential difference between them, and then recombines them in such a way that, they meet in an interference region, there will be an observable interference term in the beam intensity, i.e. $e^{i(\phi_1-\phi_2)}$, where

$$\phi_1-\phi_2 = \left(\frac{1}{\hbar}\right)\int_{t_i}^{t_f} dt\,[V_2(t)-V_1(t)]. \tag{C1}$$

One can call this as quantum interference of potentials which has been discussed in many text books, see, e.g. [50].

It is worth mentioning that the quantum interference of force is also an interesting phenomenon which has been studied recently [51].

Equation (22) has been observed and checked experimentally in physics and chemistry. One interesting recent experiment, is quantum beat spectroscopy in Helium [52]. They observed quantum beat signal given by:

$$I_{mn}(\epsilon,\tau) = A_{mp}(\epsilon,\tau)A_{np}(\epsilon,\tau)\cos\left(\omega_{mp,np}\tau+\phi_{mp,np}\right), \tag{C2}$$

which is exactly the same as the interference term in Eq. (22). Here $A_{mp}(\epsilon,\tau)$ and $A_{np}(\epsilon,\tau)$ are the ionization amplitudes from the interfering $mp$ and $np$ states and $\phi_{mp,np}$ is the phase difference between $mp$ and $np$ states due to interaction with Coulomb potential.

In [53], the authors studied the time it takes to remove an electron from an atom or molecule during photoionization. They discussed a delay time:

$$\Delta t^c = -\frac{\partial\varphi^c}{\partial I_p}, \tag{C3}$$

that comes about due to the long-range electron-core coulomb interaction. Here $\varphi^c$ is the phase accumulated by the outgoing electron due to its interaction with the ionic core and $I_p$, is the ionization potential of the bound state from which the electron escaped. $\varphi^c$ is different for different energy levels, for example for $mp$ and $np$ states.

So, the observations confirm Eqs. (20) and (22), namely the different energy states can get different additional phase, during time evolution, due to some interactions.

---

[1] R. K. Gainutdinov, J. Phys. A: Math. Gen. **22**, 269 (1989).

[2] S. R. Wilkinson *et al*., Nature **387**, 575-577 (1997).